\documentstyle[12pt,epsfig]{article}
\begin{document}
\noindent
\begin{center}
{\Large {\bf Coincidence Problem
in $f(R)$ Gravity Models}}\\ \vspace{2cm}
 ${\bf Yousef~Bisabr}$\footnote{e-mail:~y-bisabr@srttu.edu.}\\
\vspace{.5cm} {\small{Department of Physics, Shahid Rajaee Teacher
Training University,
Lavizan, Tehran 16788, Iran}}\\
\end{center}
\vspace{1cm}
\begin{abstract}
The $f(R)$ gravity models formulated in Einstein conformal frame
are equivalent to Einstein gravity together with a minimally
coupled scalar field. The scalar field couples with the matter sector and the coupling term is given by the conformal factor.
We use this interacting model to derive a necessary condition for alleviating
the coincidence problem.

\end{abstract}
%~~~~~~~PACS Numbers: 98.80.-k \vspace{3cm}
\section{Introduction}
There are strong observational evidences that the expansion of
the universe is accelerating (see e.g. \cite{super}).  However, the origin of this cosmic
acceleration is not well understood and remains as one of the main challenges of modern cosmology.  The standard explanation invokes an
unknown component, usually referred to as dark energy.  It
contributes to energy density of the universe with
$\Omega_{d}=0.7$ where $\Omega_{d}$ is the corresponding density
parameter \cite{ca}.  A
candidate for dark energy which seems to be both natural and
consistent with observations is the cosmological constant
\cite{ca} \cite{wein} \cite{cc}. However, in order to avoid
theoretical problems \cite{wein}, other scenarios have been
investigated.  In one of these scenarios the matter sector remains unchanged and
the gravitational part suffers from some modifications.   A family of these modified gravity models is obtained by
replacing
the Ricci scalar $R$ in the usual Einstein-Hilbert Lagrangian
density for some function $f(R)$ \cite{I1}.\\
There are two important problems that are related to the
cosmological constant.  The first problem, usually known as the
fine tuning problem, is the large discrepancy between observations
and theoretical predictions on its value. There have been many attempts trying to resolve this problem \cite{wein}. Most of them are based on the belief that the
cosmological constant may not have such an extremely small value
at all times and there should exist a dynamical mechanism working
during evolution of the universe which provides a cancelation of
the vacuum energy density at late times \cite{cancel}.  The second
problem concerns with the coincidence between the observed vacuum
energy density and the current matter density.  While these two
energy components evolve differently as the universe expands,
their contributions to total energy density of the universe in the
present epoch are the same order of magnitude.  Besides the possibility that the present epoch may be a stationary regime
at which the ratio of the two energy densities are constant, it is also quite possible that we live in a very special epoch, a transient epoch
at which the ratio varies slowly with respect to
the expansion of the universe.  A possible solution to the coincidence problem is to consider
an interaction between dark energy and dark matter.  If such an interaction exists the two corresponding energy densities
do not scale independently.  It is shown that, this can lead to a constant ratio of energy densities when an appropriate
coupling term is applied \cite{c2} \cite{c3}.\\
In the present note, we will consider the coincidence problem in Einstein frame representation of $f(R)$ gravity models.
In these models the dynamical variable of the vacuum sector
is the metric tensor and the corresponding field equations are fourth order.  This
dynamical variable can be replaced
by a new pair which consists of a conformally rescaled metric and a scalar partner.  Moreover, in terms of the new set
of variables the field equations are those of General Relativity.  The original set of variables
is commonly called Jordan conformal frame and the transformed set whose dynamics is described by Einstein field equations
is called Einstein conformal frame. The dynamical
equivalence of Jordan and Einstein conformal frames does not
generally imply that they are also physically equivalent.  In fact
it is shown that some physical systems can be differently
interpreted in different conformal frames \cite {soko} \cite{no}.
The physical status of the two conformal frames is an open
question which we are not going to address here.  Our motivation to work in Einstein conformal frame is that in this frame there is a coupling between the scalar degree of freedom and matter
sector induced by the conformal transformation.  As previously stated, there is a large amount of interest to realize the coincidence
problem as a consequence of an interaction between matter systems and the dark sector.  Although the whole idea seems to be promising, however, the
suggested interaction terms are usually phenomenological and are not generated by a fundamental theory.  In our case the interaction term is
given by the conformal factor.  We investigate the consequences of this interaction term and derive an expression which constrains the form of the $f(R)$ function.  We will show that this constraint selects those $f(R)$ models that allow for possible alleviation of the coincidence problem.
~~~~~~~~~~~~~~~~~~~~~~~~~~~~~~~~~~~~~~~~~~~~~~~~~~~~~~~~~~~~~~~~~~~~~~~~~~~~~~~~~~~~~~~~~~

\section{Framework}
 The action for an $f(R)$
gravity theory in the Jordan frame is given by
\begin{equation}
S_{JF}= \frac{1}{2k}\int d^{4}x \sqrt{-g} f(R) +S_{m}(g_{\mu\nu}, \psi)\label{b1}\end{equation} where $k\equiv 8\pi G$, $G$ is
the gravitational constant, $g$ is the
determinant of $g_{\mu\nu}$ and $S_{m}$ is the action
of (dark) matter which depends on the metric $g_{\mu\nu}$ and some (dark) matter
field $\psi$.  Stability in matter sector (the Dolgov-Kawasaki
instability \cite{dk}) imposes some conditions on the functional
form of $f(R)$ models.  These
conditions require that the first and the second derivatives of
$f(R)$ function with respect to the Ricci scalar $R$ should be
positive definite.  The positivity of the first derivative ensures
that the scalar degree of freedom is not tachyonic and positivity
of the second derivative tells us that graviton is not a ghost.\\
It is well-known that $f(R)$ models are equivalent to
a scalar field minimally coupled to gravity with an appropriate
potential function.  In fact, we may use a new set of variables
\begin{equation}
\bar{g}_{\mu\nu} =\Omega~ g_{\mu\nu} \label{b2}\end{equation}
\begin{equation} \phi = \frac{1}{2\beta \sqrt{k}} \ln \Omega
\label{b3}\end{equation}
 where
$\Omega\equiv\frac{df}{dR}=f^{'}(R)$ and
$\beta=\sqrt{\frac{1}{6}}$. This is indeed a conformal
transformation which transforms the above action in the Jordan
frame to the following action in the Einstein frame \cite{soko} \cite{w}
\begin{equation}
S_{EF}=\frac{1}{2} \int d^{4}x \sqrt{-\bar{g}}~\{\frac{1}{k}
\bar{R}-\bar{g}^{\mu\nu} \nabla_{\mu} \phi~ \nabla_{\nu} \phi
-2V(\phi)\}+ S_{m}(\bar{g}_{\mu\nu}e^{2\beta \sqrt{k}\phi}
, \psi) \label{b4}\end{equation}All indices are raised and lowered by $\bar{g}_{\mu\nu}$ unless
stated otherwise.  In the Einstein frame, $\phi$ is a minimally
coupled scalar field with a self-interacting potential which is
given by
\begin{equation}
V(\phi(R))=\frac{Rf'(R)-f(R)}{2kf'^2(R)}
\label{b5}\end{equation} Note that the conformal transformation
induces the coupling of the scalar field $\phi$ with the matter
sector. The strength of this coupling $\beta$, is fixed to be
$\sqrt{\frac{1}{6}}$ and is the same for all types of matter
fields.  In the action (\ref{b4}), we take $\bar{g}^{\mu\nu}$ and
$\phi$ as two independent field variables and variations of the
action yield the corresponding dynamical field equations.
Variation with respect to the metric tensor $\bar{g}^{\mu\nu}$,
leads to
\begin{equation}
\bar{G}_{\mu\nu}=k(\bar{T}^{\phi}_{\mu\nu}+
\bar{T}^{m}_{\mu\nu}) \label{b6}
\label{b7}\end{equation} where
\begin{equation}
\bar{T}^{\phi}_{\mu\nu}=\nabla_{\mu}\phi
\nabla_{\nu}\phi-\frac{1}{2}\bar{g}_{\mu\nu}\nabla^{\gamma}\phi
\nabla_{\gamma}\phi-V(\phi)\bar{g}_{\mu\nu}
\label{b8}\end{equation}
\begin{equation}
\bar{T}^m_{\mu\nu}=\frac{-2}{\sqrt{-\bar{g}}}\frac{\delta S_{m}(\bar{g}_{\mu\nu}, \psi)}{\delta \bar{g}^{\mu\nu}} \label{b9}\end{equation} are
stress-tensors of the scalar field and the matter field system.
The trace of (\ref{b7}) is
\begin{equation}
\nabla^{\gamma}\phi
\nabla_{\gamma}\phi+4V(\phi)-\bar{R}/k=\bar{T}^m\label{b8-1}\end{equation}
which differentially relates the trace of the matter stress-tensor $\bar{T}^{m}=\bar{g}^{\mu\nu}\bar{T}^m_{\mu\nu}$
to $\bar{R}$.  Variation of the action
(\ref{b4}) with respect to the scalar field $\phi$, gives
\begin{equation}
\bar{\Box}\phi-\frac{dV(\phi)}{d\phi}=-\beta \sqrt{k}\bar{T}^{m}
\label{b11}\end{equation}
It is important to note that the two
stress-tensors $\bar{T}^m_{\mu\nu}$ and $\bar{T}^{\phi}_{\mu\nu}$
are not separately conserved.
 Instead they satisfy the following equations
\begin{equation}
\bar{\nabla}^{\mu}\bar{T}^{m}_{\mu\nu}=-\bar{\nabla}^{\mu}\bar{T}^{\phi}_{\mu\nu}= \beta \sqrt{k}\nabla_{\nu}\phi~\bar{T}^{m}\label{b13}\end{equation} We apply the field equations in a
spatially flat homogeneous and isotropic cosmology described by
Friedmann-Robertson-Walker spacetime
\begin{equation}
ds^2=-dt^2+a^2(t)(dx^2+dy^2+dz^2)
\end{equation}
where $a(t)$ is the scale factor. To do this, we take
$\bar{T}^m_{\mu\nu}$ and $\bar{T}^{\phi}_{\mu\nu}$ as the stress-tensors of a pressureless perfect fluid with energy density
$\bar{\rho}_{m}$, and a perfect fluid with energy density
$\rho_{\phi}=\frac{1}{2}\dot{\phi}^2+V(\phi)$ and pressure
$p_{\phi}=\frac{1}{2}\dot{\phi}^2-V(\phi)$, respectively. In this
case, (\ref{b7}) and (\ref{b11}) take the form \footnote{Hereafter we will use unbarred characters in the Einstein frame.}
\begin{equation}
3H^2=k(\rho_{\phi}+\rho_{m})
\label{b14}\end{equation}
\begin{equation}
2\dot{H}+3H^2=-k\omega_{\phi}\rho_{\phi}
\label{b14-1}\end{equation}
\begin{equation}
\ddot{\phi}+3H\dot{\phi}+\frac{dV(\phi)}{d\phi}=-\beta \sqrt{k}\rho_{m}
\label{b15}\end{equation} where
$\omega_{\phi}=\frac{p_{\phi}}{\rho_{\phi}}$ is equation of state parameter of the scalar field $\phi$, and overdot indicates differentiation with respect
to cosmic time $t$.  The trace equation (\ref{b8-1}) and the conservation equations
(\ref{b13}) give, respectively,
\begin{equation}
\dot{\phi}^2+R/k-4V(\phi)=\rho_{m}
\label{b16}\end{equation}
\begin{equation}
\dot{\rho}_{m}+3H\rho_{m}=Q \label{b17}\end{equation}
\begin{equation}
\dot{\rho}_{\phi}+3H(\omega_{\phi}+1)\rho_{\phi}=-Q
\label{b18}\end{equation} where
\begin{equation}
Q=\beta \sqrt{k}\dot{\phi}\rho_{m}
\label{b-18}\end{equation} is the interaction term.  This term
vanishes only for $\phi=const.$, which due to (\ref{b3}) happens
when $f(R)$ linearly depends on $R$. The direction of energy
transfer depends on the sign of $Q$ or $\dot{\phi}$.  For
$\dot{\phi}>0$, the energy transfer is from dark energy to dark
matter and for $\dot{\phi}<0$ the reverse is true.\\  We emphasize that the coupling term (\ref{b-18}) is very similar to some phenomenological
 coupling terms suggested in the literature.  In fact, there are different kinds of interacting models which have been investigated \cite{c2} \cite{c3}.  A particular class of these models considers $Q=\alpha \dot{\varphi} \rho$ in which $\alpha$ is a coupling constant, $\varphi$ is usually a quintessence field and $\rho$ is energy density of dark matter \cite{c3}. Apart from the similarity of the latter with (\ref{b-18}), there are also some important differences.  Firstly, the scalar field $\phi$ is not a kind of matter field and is actually given in terms of the function $f(R)$.  Secondly, $\beta$ is a universal coupling constant implying that $\phi$ couples with the same strength to all types of matter fields.  In contrary,  it is possible to consider $\alpha$ as a non-universal coupling constant so that it may couple to dark matter and baryons with different strengths \cite{dar}.  Moreover, the value of $\beta$ is fixed to be $1/\sqrt{6}$ while $\alpha$ is constrained by observations \cite{c3}.  We will return to this last point later.
~~~~~~~~~~~~~~~~~~~~~~~~~~~~~~~~~~~~~~~~~~~~~~~~~~~~~~~~~~~~~~~~~~~~~~~~~~~~~~~~
\section{The Coincidence Problem}
One of the important features of the cosmological constant problem
is the present coincidence between dark energy and dark matter
energy densities \cite{c1}.  There is a class of models in which
this observation is related to some kinds of interaction between the
two components \cite{c2} \cite{c3}.  In these models the two components are not
separately conserved and there is a flow of energy from dark
energy to dark matter or vice versa. In this sense,
dark energy and dark matter energy densities may
have the same scaling at late times due to the interaction, although they decrease with the
expansion of the universe at different rates.
 The important task in this context is
to find a constant ratio of dark energy to dark matter energy densities
for an appropriate interaction term.  Despite the fact that this approach seems to
be promising, there is not still a
compelling form of interaction which is introduced by a
fundamental theory.  Therefore one usually uses different
interaction terms and tries to adapt them with recent observations.\\
In $f(R)$ gravity models presented in the Einstein frame, there is
a fixed interaction between the scalar field and matter sector.  Since the form of the interaction is fixed by the conformal transformation
one can therefore search for some appropriate forms of the function $f(R)$ for
 which the energy densities ratio of the two components takes a stationary value. This is the strategy that we are going to
pursue in this section, namely, to find some conditions on the
functional
form of $f(R)$ that may lead to a constant $r\equiv \rho_{m}/\rho_{\phi}$ .\\
To do this, we consider time evolution of the ratio $r$ ,
\begin{equation}
\dot{r}=\frac{\dot{\rho}_{m}}{\rho_{\phi}}-r\frac{\dot{\rho}_{\phi}}{\rho_{\phi}}
\label{c1}\end{equation} From equations (\ref{b17}), (\ref{b18}) and
(\ref{b-18}) we obtain
\begin{equation}
\dot{r}=3Hr
\omega_{\phi}+\beta \sqrt{k}\dot{\phi}r(r+1)
\label{c2}\end{equation} In this relation, we can write $\dot{r}$ in terms of the parameters $r$ and $q$.  We first use (\ref{b14}) and
(\ref{b14-1}) to replace the equation of state parameter $\omega_{\phi}$
with the deceleration parameter $q$.  Applying
\begin{equation}
\dot{H}=-(q+1)H^2 \label{c3}\end{equation} to the equation
(\ref{b14-1}) gives
\begin{equation}
\omega_{\phi}=\frac{(2q-1)H^2 }{\sqrt{k}\rho_{\phi}}
\label{c4}\end{equation}
We then use (\ref{b14}) in the latter and substitute the result in (\ref{c2}), which leads to
\begin{equation}
\dot{r}=Hr(2q-1)(r+1)+\beta \sqrt{k}\dot{\phi}r(r+1)
\label{c5}\end{equation}
On the other hand, we can combine the trace equation (\ref{b16}) with the equations (\ref{b5}) and (\ref{b14}) to obtain
\begin{equation}
\dot{\phi}^2=\frac{1}{k}\{\frac{3H^2 r}{r+1}+3H^2(2q-3)(1-\frac{2}{f'}) -2\frac{f}{f'^{2}}\}
\label{c6}\end{equation}
When we put this expression into (\ref{c5}), the result is an equation that relates $\dot{r}$ to the parameters $r$, $q$ and $H$.  The
requirement that the universe
approach a stationary stage in which $r$ either becomes a constant or varies more slowly than the scale factor, leads to the following relation
\begin{equation}
g(f'; H, r_{s}, q)=0
\label{c7-1}\end{equation}
where
\begin{equation}
g(f'; H, r_{s}, q)\equiv r_{s}(2q-1)(r_{s}+1)+\beta r_{s}(r_{s}+1)\{\frac{3r_{s} }{r_{s}+1}+3(2q-3)(1-\frac{2}{f'})
-\frac{2f}{H^2 f'^{2}}\}^{\frac{
1}{2}}
\label{c7}\end{equation}
and $r_{s}$ is the value of $r$ when it takes a stationary value.
It is now possible to use (\ref{c7-1}) to check that whether a particular $f(R)$ model
is consistent with a late-time stationary ratio of energy densities. In general, to find
such $f(R)$ gravity models one may start with a particular $f(R)$
function in the action (\ref{b1}) and solve the corresponding
field equations for finding the form of $q(z)$ or $H(z)$.
However, this approach is not efficient in view of
complexity of the field equations.  An alternative
approach is to start from the best fit parametrization $q(z)$
obtained directly from data and use this $q(z)$ for a particular
$f(R)$ function in (\ref{c7-1}).  Here we will follow the latter approach. \\For a given redshift $z_{0}$ and the parameters $r_{s}(z_{0})$, $q(z_{0})$
and $H(z_{0})$, the relation (\ref{c7-1}) acts as a constraint on the function $f(R)$.
As an illustration, we apply this constraint to some $f(R)$ functions. Before doing this, there
are some remarks to do with respect to (\ref{c7-1}). This condition is a consequence of $\dot{r}=0$ when $r=r_{s}$ becomes stationary at late-times.  At sufficiently late-times characterized by $z=z_{0}$, we take $r_{s}=r_{0}$ and rewrite (\ref{c7-1}) as
\begin{equation}
g(f'_{0}; H_{0}, r_{0}, q_{0})=0
\label{d10-3}\end{equation}
where
\begin{equation}
g(f'_{0}; H_{0}, r_{0}, q_{0})\equiv r_{0}(2q_{0}-1)(r_{0}+1)+\beta r_{0}(r_{0}+1)\{\frac{3r_{0} }{r_{0}+1}+3(2q_{0}-3)(1-\frac{2}{f_{0}'})
-2\frac{f_{0}}{H_{0} f_{0}'^{2}}\}^{\frac{
1}{2}}
\label{d10-4}\end{equation}
Here the functions $f_{0}$, $f'_{0}$ and $f''_{0}$ are
the late-time configurations of $f(R)$, $f'(R)$ and $f''(R)$ which are obtained by replacing $R$ with
\begin{equation}
R=6(1-q)H^2 \label{d10-2}\end{equation}
at the redshift $z_{0}$.  Note that
an $f(R)$ gravity model is usually given in terms of some parameterizations. In this sense, the condition (\ref{c7-1})
acts actually as a constraint relating the corresponding parameters of a particular $f(R)$ gravity model to the constants $q_{0}$, $r_{0}$ and $H_{0}$.\\
We use a two-parametric reconstruction function for characterizing $q(z)$ \cite{wang}\cite{q},
\begin{equation}
q(z)=\frac{1}{2}+\frac{q_{1}z+q_{2}}{(1+z)^2}
\label{d10-1}\end{equation} Fitting this model to the Gold
data set gives $q_{1}=1.47^{+1.89}_{-1.82}$ and $q_{2}=-1.46\pm
0.43$ \cite{q}.  We also take $z_{0}=0.25$ which, with use of (\ref{d10-1}), corresponds to $q_{0}\approx -0.2$.  Moreover, recent
observations imply that $r_{0}\equiv \frac{\rho_{m}(z_{0})}{\rho_{\phi}(z_{0})}\approx \frac{3}{7}$ \cite{dd}.\\  Now let
us first consider the
model \cite{cap} \cite{A}
\begin{equation}
f(R)=R+\lambda R_{0} (\frac{R}{R_{0}})^n \label{d11}\end{equation}
Here $R_{0}$ is taken to be of the order of $H_{0}^2$ and $\lambda$, $n$ are constant parameters. In terms of the values
attributed to these parameters, the model
(\ref{d11}) is divided by three cases \cite{A}. Firstly, when
$n>1$ there is a stable matter-dominated era which does not follow
by an asymptotically accelerated regime. In this case, $n = 2$
corresponds to Starobinsky's inflation and the accelerated phase
exists in the asymptotic past rather than in the future. Secondly,
when $0<n<1$ there is a stable matter-dominated era followed by an
accelerated phase only for $\lambda<0$. Finally, in the case that
$n<0$ there is no accelerated and matter-dominated phases for
$\lambda>0$ and $\lambda<0$, respectively.  Thus the model
(\ref{d11}) is cosmologically viable in the regions of the
parameters space which is given by $\lambda<0$ and $0<n<1$.\\
When we use (\ref{d10-2}) in the function
$g(f_{0}'; H_{0}, r_{0}, q_{0})$, it takes the form of an expression which relates the parameters $n$ and $\lambda$ to $q_{0}$, $r_{0}$ and $H_{0}$.  In
fig.1 we have plotted $g(n, \lambda; H_{0}, r_{0}, q_{0})$ for $\lambda=-1$.
\begin{figure}[ht]
\begin{center}
\includegraphics[width=0.45\linewidth]{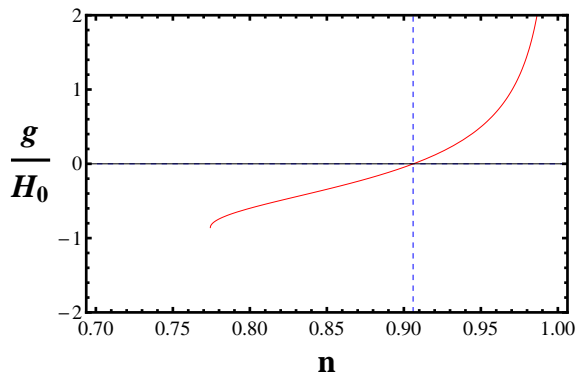}
\caption{ The plot of $g(n, \lambda; H_{0}, r_{0}, q_{0})$ for
the model (\ref{d11}) when $\lambda =-1$, $q_{0}=-0.2$ and $r_{0}=3/7$.  The vertical dashed line corresponds to $n=0.906$.  }
\end{center}
\end{figure}
This figure indicates that the constraint (\ref{d10-3}) is satisfied only for $n\approx 0.9$ which implies that for this value of the parameter $n$, the
 model (\ref{d11}) admits a late-time stationary ratio of the energy densities.   Note that $n\approx 0.9$ lies in the range that the model
 is cosmologically viable. \\
Now we consider the model presented by Starobinsky \cite{star} \cite{sta}
\begin{equation}
f(R)=R-\gamma R_{0} \{1-[1+(\frac{R}{R_{0}})^2]^{-m}\}
\label{d12}\end{equation}
where $\gamma$, $m$ are
positive constants and $R_{0}$ is again of the order of the
presently observed effective cosmological constant. Using the
same procedure, we have
plotted the function $g(m, \gamma; H_{0}, r_{0}, q_{0})$ in fig.2.
\begin{figure}[ht]
\begin{center}
\includegraphics[width=0.42\linewidth]{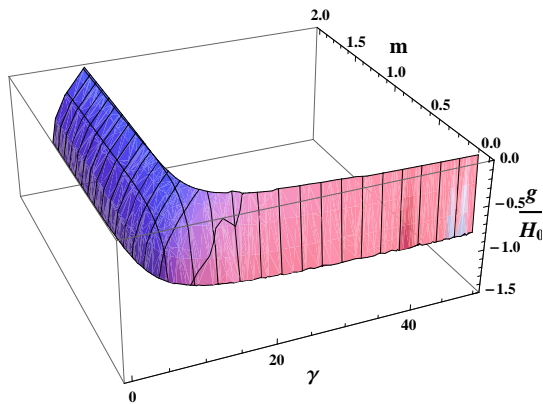}
\caption{ The plot of $g(m, \gamma; H_{0}, r_{0}, q_{0})$ for
the model (\ref{d12}) when $q_{0}=-0.2$ and $r_{0}=3/7$.}
\end{center}
\end{figure}
The figure shows that there are some regions in the parameters space for which the condition (\ref{d10-3}) is
satisfied.  The condition is satisfied on the upper boundary of the surface plotted in fig.2 where
$g(m, \gamma; H_{0}, r_{0}, q_{0})=0$.  Thus for
the corresponding values of the parameters, the coincidence problem can be addressed in the context of the model (\ref{d12}).
    For instance, as the figure indicates the parameters space is bounded by $\gamma \geq 10.5$ and $m \geq 0.04$ so that for $m>0.04$ the parameter $\gamma$ should remain near the value $10.5$.

~~~~~~~~~~~~~~~~~~~~~~~~~~~~~~~~~~~~~~~~~~~~~~~~~~~~~~~~~~~~~~~~~~~~~~~~~~~~~~~~~~~~~~~~~~
\section{Conclusion}
In Einstein frame representation of $f(R)$ gravity models, the scalar partner of the metric tensor
interacts with (dark) matter in such a way that the interaction term is fixed by the conformal transformation.  This means that contributions of the scalar field and the (dark) matter system to total energy density do not scale independently.  As a consequence, even tough the two components
may start with different scalings at early times, they may have the same scaling at sufficiently late times. \\
   We have considered this feature as a possibility for addressing the coincidence problem.  In fact, the interaction of dark energy and dark matter has been recently taken as a natural guidance for alleviating the coincidence problem by some authors.  In absence of an interaction or coupling term based on a fundamental theory, most of the current investigations have been limited to a phenomenological level.  In our case, the interaction term, $Q$, is given by the conformal transformation and can be written in terms of $\dot{R}$, $f'(R)$ and $f''(R)$.  Due to stability considerations, any viable $f(R)$ model should satisfy $f'(R)>0$ and $f''(R)>0$ \cite{do}.  Thus the direction of the energy transfer is determined by the sign of $\dot{R}$ in a particular epoch.  For instance, in an epoch for which $\dot{R}>0$, the energy transfer is from dark energy (or the scalar field $\phi$) to dark matter while for $\dot{R}<0$ the reverse is true.\\
      We  have derived a relation giving the evolution of the parameter $r$. We have found that there is a class of $f(R)$ gravity models satisfying the condition (\ref{d10-3}) for which a late-time stationary state for $r$ exists.  As illustrations, we have shown that the model (\ref{d11}) lies in this class only for $n \approx 0.9$.  The condition is also used for the Starobinsky's model.   We have shown that there is a region in the parameters space for which the coincidence problem can be addressed in this model.  The region is characterized by the upper border of the surface plot of the fig.2 for which $g(m, \gamma; H_{0}, r_{0}, q_{0})=0$.\\ Finally, we point out that there is no a free parameter in the interaction term (\ref{b-18}) since $\beta$ is fixed by conformal transformation. In general, the interaction of the scalar field $\phi$ and the matter sector may lead to a fifth force and violation of equivalence principle.  In fact, the real challenge for alleviating the
coincidence problem comes from the combination of restrictions from
local gravity experiments and dynamical considerations.
       Thus the question is that how a coupling term without a free parameter can be consistent with local gravity experiments.  The point is that, in our case, these experiments constrain the corresponding parameters of a particular $f(R)$ gravity model\footnote{These constraints are imposed by chameleon mechanism. See, e.g., \cite{bis} and references therein.} rather than the coupling constant of the interaction term.  For the model (\ref{d12}), it is shown \cite{capo} that the most stringent bound is $m>0.9$ which comes from violation of equivalence principle.  Combining the latter with the bounds indicated in fig.2, one infers that alleviation of the coincidence problem requires that $\gamma \approx 10.5$.

\end{document}